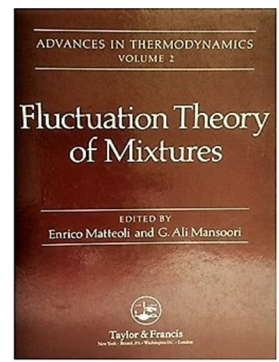

# Fluctuation Theory of Mixtures
## (A Statistical Mechanical Background)

**G. Ali Mansoori**
*University of Illinois at Chicago (M/C 063), Chicago, IL 60607-7052 USA, mansoori@uic.edu*

**Enrico Matteoli**
*Istituto di Chimica Quantistica ed Energetica Molecolare del CNR, Via Risorgimento 35 56126 Pisa, Italy, matteoli@ipcf.cnr.it*

## ABSTRACT

The statistical mechanical basis of the fluctuation theory of mixtures is reviewed. An overview of the statistical mechanical relations between the microscopic properties of a system and its macroscopic properties is presented. The distribution functions in equilibrium systems are defined and the relations between distribution functions and thermodynamic properties for pure fluids and fluid mixtures are reported. The available approximations to distribution functions are presented. Definitions of the direct correlation functions, direct correlation function integrals, and their relations with fluctuation integrals are reported.

## INTRODUCTION

Scientists in their efforts to develop new ideas and engineers in their endeavors to build and improve industrial processes and efficient utilization of the available resources, deal with variety of physical, chemical, and biological phenomena. This requires detailed knowledge of the thermodynamic and transport properties of the substances being considered. In thermodynamics, the effects of pressure, temperature, and mixing with other substances on thermodynamic properties are of major concern.

Experimental measurements of thermodynamic properties can not cover all the needs of science and industry. They are costly and time consuming. The large number of substances and the wide range of conditions we have to deal with make it rather difficult to rely completely on experimental measurements. The problem becomes more complicated for mixtures, because the number of measurements needed grows rapidly with the number of components in a given mixture. Because of this problem, it becomes necessary to minimize the amount of experimental measurements by developing theoretical models that could be capable of correlating the existing data and extrapolating to regions where none or little experimental data are available.

The early work in this field relied mainly on empirical approaches. For example, Kamerlingh Onnes (1901) developed the empirical version of the virial equation of state. Later, it was shown that this equation had a solid theoretical basis. The van der Waals equation (1873) was derived from intuitive ideas. This equation was then proved to be theoretically sound (Prins, 1974). It has been also modified, both theoretically and empirically, by many workers.

The theoretical tools needed to describe the behavior of real fluids at equilibrium were developed in the late nineteenth century by Boltzmann and Gibbs. They laid the foundation for statistical mechanics. However, the astronomical amount of calculations needed to get the thermodynamic properties from molecular data delayed the use of statistical mechanics in practical problems for some time. The work of Kirkwood and others in the early part of the twentieth century opened the door for applying statistical mechanics in practical thermodynamic problems (Haile and Mansoori, 1983).

In their effort to model the behavior of real mixtures, scientists and engineers follow two methods: activity coefficient models, and equations of state approach. The early models of the activity coefficients were based on composition power series expansions of the excess Gibbs free energy. Margules and van Laar equations (Walas, 1985) are examples of such expansions. Local composition models have been used to obtain activity coefficient expressions by a number of workers (Wilson, 1964; Renon and Prausnitz, 1968). The connection between local composition and statistical mechanics was later shown in a number of ways (Kemeny and Rasmussen, 1981; Hu et al., 1983; Mansoori and Ely, 1985b). Abrams and Prausnitz (1975) applied statistical mechanics directly to obtain expressions for the activity coefficients.

Statistical mechanical equations of state can be obtained in a number of ways. One route is to make an assumption that allows evaluating analytically the integrals in the partition function. The lattice model does this by assuming that the molecules can move in only defined volumes around their equilibrium positions. This assumption makes the integrals in the partition function independent and as a result each integral can be evaluated separately. This approach, however, is successful in describing the behavior of solids but not fluids.

Another approach to obtaining equations of state from statistical mechanics is to assume the partition function to be divided into repulsive and attractive parts. This implies that the same division holds for the equation of state. Usually the repulsive part is taken as that of a hard body, where the potential is infinite inside the core of the particles and zero outside. Accurate statistical mechanical expressions for the properties of such particle systems are available for both pure fluids and mixtures. The attractive part is usually treated as a perturbation expansion around the repulsive part.

The distribution function approach to the equilibrium properties of substances relates the relative locations of particles in a given substance to the thermodynamic properties of the substance. The distribution functions are normalized probability densities for the locations of n particles at given intermolecular distances from a central particle. Information about the distribution functions of particles in real fluids are not quite available. As a result, approximations to the distribution functions are used to get useful practical results. In the perturbation and the variational theories the distribution functions of a reference fluid (usually a hard-sphere fluid) are used rather than the real functions. These two theories usually require the knowledge of the potential energy functions between the molecules. This limits their use to simple fluids for which intermolecular potential energy functions and data are available.

Modeling of the behavior of complex mixtures can be accomplished more effectively provided one assumes the properties of the pure components are known. A large amount of experimental data has been collected for the widely used pure fluids, and has been put into the format of empirical equations of state. The contribution of statistical mechanics can be more effective in predicting mixture properties from the existing pure fluid equations of state.

## CLASSICAL AND STATISTICAL THERMODYNAMICS

Collection of particles (atoms or molecules) possess both microscopic and macroscopic properties. The thermodynamic properties, such as pressure, internal energy and entropy are macroscopic properties. Through classical thermodynamics we can derive relations which allow calculating one thermodynamic property from another. Such relations have the advantage of being applicable to all kinds of systems at equilibrium. Techniques of statistical mechanics are the needed tools for predicting the behavior of a given substance from microscopic information or microscopic energy levels. The energy levels, in turn, can be obtained from the knowledge of the potential energy between the particles forming the system and from the system volume. The energy levels are obtained from quantum mechanics, usually through the Schrödinger equation. The energy levels are the eigenvalues of this equation. However, it is usually difficult to obtain a solution to the Schrödinger equation. In fact the Schrödinger equation has not been solved exactly for systems consisting of more than two particles, except when the potential energy function is assumed to be zero everywhere. In real systems the number of particles is of the order of $10^{23}$. Such a large number of interacting particles makes it extremely difficult to obtain the quantum mechanical energy levels.

Assuming the energy levels are available, the macroscopic properties can then be obtained from the partition function. In the canonical ensemble where the temperature, the volume and the number of particles of each component in the system, $N_i$, are fixed, the partition function, Q, for a c-component system is:

$$Q(T,V,N) = \sum e^{-\varepsilon_j/kT}, \qquad (1)$$

where k is Boltzmann constant, T is the absolute temperature, V is the system volume, $\varepsilon_j$ is the energy of the particles at level j and $N=N_1, N_2, ....., N_c$. The summation in Equation (1) is carried over all the molecules in all energy levels. The canonical partition function, Q, is related to the macroscopic Helmholtz free energy of the system, A, by

$$A = -kT \ln Q(T,V,N). \qquad (2)$$

The relation between the microscopic and the macroscopic properties can be obtained for other kinds of ensembles as well. For example, in the grand canonical ensemble, where the temperature, the volume and the chemical potential of each component in the system are fixed, the partition function, $\Xi$ is:

$$\Xi(T,V,\mu) = \sum_{N_1} \cdots \sum_{N_c} Q(T,V,N) \, e^{\mu.N/kT}, \qquad (3)$$

where $N=\mu_1, \mu_2, ..., \mu_c$ and the scalar product $\mu.N$ is:

$$\mu.N = \sum_{k=1}^{c} \mu_k N_k. \qquad (4)$$

The relation between the grand canonical ensemble partition function and the system pressure, P, is

$$PV = kT \ln \Xi(T,V,\mu). \qquad (5)$$

Equations (1) and (3) are completely general for equilibrium systems. However, as mentioned above it is very difficult to get the quantum mechanical energy levels for real systems. To avoid this problem, one makes use of the observation that the spacings between the translational energy levels for most molecules away from absolute zero are very small. This means that one can treat the translational energy of the molecules as continuous rather than discrete without introducing a detectable error. Since this is not generally the case for the rotational and vibrational energy levels for molecules with internal motion, it is advantageous to separate the contributions of these two motions from the contribution of the translational motion.

To achieve the separation between internal and translational motions it is usually assumed that the internal motion of a molecule does not interfere appreciably with the motion of the center of its mass. The partition function separates, then, into two parts. The first part, which is called the configurational partition function, represents the motion of the centers of mass in the potential field created by the interaction between molecules. The second part, which is called the internal partition function, represents the contribution from the internal motions of the molecules and it is equal to the internal partition function of an ideal gas under the same conditions. The small energy gap between consecutive energy levels for all molecules, except the lightest, such as $H_2$ and He, allows replacing the summation in Equation (1) by integration over the phase space. Thus obtaining the semi-classical partition function as:

$$Q(T,V,N) = Q_{int} \prod_i^c (N_i! \lambda_i^{3N_i})^{-1} \int_V \ldots \int_V e^{-\Phi/kT} dr_1 \ldots dr_N, \qquad (6)$$

where $Q_{int}$ is the internal partition function, $\prod$ denotes the product over i, $\lambda_i$ is the De Broglie thermal wavelength of component i,

$$\lambda_i = h/(2\pi m_i kT), \qquad (7)$$

h is Planck's constant, $m_i$ is the mass of molecule i, $\Phi$ is the total potential energy of the system and $r_i$ is the position vector of molecule i. All the integrals in Equation (6) are carried over the system volume.

## DISTRIBUTION FUNCTIONS

The partition function can be evaluated, at least in principle, by carrying out the integrations in Equation (6) for a substance with known potential function. However, this task is rather difficult because of the very large number of molecules involved in real systems. A more convenient statistical mechanical formulation is based on the concept of distribution functions. The probability, $P^{(N)}$, of finding molecule 1 in volume element $dr_1$ at $r_1$, molecule 2 in volume element $dr_2$ at $r_2, \ldots,$ and molecule N in volume element $dr_N$ at $r_N$ is given by (Hill, 1956)

$$P^{(N)} dr_1 \ldots dr_N = e^{-\Phi/kT} dr_1 \ldots dr_N / Z(T,V,N), \qquad (8)$$

where $Z(T,V,N)$ is the configurational integral,

$$Z(T,V,N) = \int \ldots \int e^{-\Phi/kT} dr_1 \ldots dr_N. \qquad (9)$$

Usually we are interested in the relative position of two molecules, irrespective of the location of the other molecules in the system. This can be obtained by integrating Equation (8) over the positions of all molecules except those which we are interested in. This leads to the definition of the distribution function, $\rho^{(2)}_{ij}(r_1,r_2)$, which gives the probability of finding a molecule of type i in $dr_1$ at $r_1$ and molecule of type j in $dr_2$ at $r_2$,

$$\rho^{(2)}_{ij}(r_1,r_2) = N_i(N_j - \delta_{ij}) \int \ldots \int e^{-\Phi/kT} dr_3 \ldots dr_N / Z(T,V,N), \qquad (10)$$

where $\delta_{ij}$ is the Kronecker delta. Note that $\rho^{(2)}_{ij}(r_1,r_2)$ depends on temperature, density, and composition in addition to $r_1$ and $r_2$. For molecules which interact with radially symmetric potential functions $\rho^{(2)}_{ij}$ depends only on the distance between the centers of masses,

$$\rho^{(2)}_{ij}(r_1,r_2) = \rho^{(2)}_{ij}(r_{12}); \quad r_{12} = |r_1 - r_2|.$$

In the limit of ideal gas ($\Phi/kT \to 0$) the distribution function $\rho^{(2)}_{ij}(r_1,r_2)$ approaches the value $N_i(N_j - \delta_{ij})/V^2$. This suggests defining the pair (or radial) distribution function, $g_{ij}(r)$, by

$$g_{ij}(r) = \rho^{(2)}_{ij}(r) V^2 / (N_i N_j), \qquad (11)$$

which approaches $1 - \delta_{ij}/N_j$ in the ideal gas limit. Combining Equations (10) and (11) gives

$$g_{ij}(r) = V^2 (1 - \delta_{ij}/N_j) \int \ldots \int e^{-\Phi/kT} dr_3 \ldots dr_N / Z(T,V,N). \qquad (12)$$

A similar definition for the distribution function exists in the grand canonical ensemble as the following

$$g_{ij}(r) = V^2/(N_i N_j) \sum_{N_i \geq i} \sum_{N_j \geq j} \cdots \sum_{N_c}^{c} \pi(N!_k) e^{\mu \cdot N/kT} \int \cdots \int e^{-\Phi/kT} dr_3 \cdots dr_N / \Xi(T,V,\mu) \quad (13)$$

The pair (or radial) distribution function in the grand canonical ensemble approaches unity in the limit of ideal gas.

## RELATIONS BETWEEN THERMODYNAMIC PROPERTIES AND g(r)

The relation between $g(r)$ and the internal energy, $U$, can be derived from the definition of the internal energy in terms of the canonical ensemble partition function:

$$U - U^\circ = kT^2 [\partial \ln Z(T,V,N)/\partial T]_{N,V}, \quad (14)$$

where $U^\circ$ is the ideal gas internal energy. Substituting for $Z(T,V,N)$ from Equation (9) we get

$$U - U^\circ = \int \cdots \int e^{-\Phi/kT} dr_1 \cdots dr_N / Z(T,V,N). \quad (15)$$

For pure fluids under the assumption of pairwise additivity of intermolecular interactions one can write $\Phi$ as the summation of $N(N-1)/2$ potential energy terms

$$\Phi = \sum_{i=1}^{N} \sum_{j<i}^{N} \phi(r_{ij}). \quad (16)$$

Upon substitution of Equation (16) into Equation (15) the relation between the internal energy and the radial distribution function for pure fluids is obtained:

$$U - U^\circ = 1/2 \, N \, \rho \int_0^\infty \phi(r) \, g(r) \, 4\pi r^2 \, dr, \quad (17)$$

where $\rho$ is the molecular number density, $N/V$. Equation (17) can also be derived from the grand canonical ensemble distribution function. For mixtures of c-components Equation (17) takes the form

$$U - U^\circ = 1/2 \, N \, \rho \sum_{i=1}^{c} \sum_{j=1}^{c} x_i \, x_j \int_0^\infty \phi_{ij}(r) \, g_{ij}(r) \, 4\pi r^2 \, dr, \quad (18)$$

where $x_i$ is the mole fraction of component $i$ and $\phi_{ij}$ is the pair intermolecular potential energy function for molecules of type i and j. For pure fluids the pressure is given in terms of g(r) by (Hill, 1956)

$$P = kT\rho - \rho^2/6 \int_0^\infty r \phi'(r) g(r) \, 4\pi r^2 \, dr. \tag{19}$$

For mixtures, Equation (18) takes the following form

$$P = kT\rho - \rho^2/6 \sum_{i=1}^{c} \sum_{j=1}^{c} x_i x_j \int_0^\infty r \phi'_{ij}(r) g_{ij}(r) \, 4\pi r^2 \, dr, \tag{20}$$

where $\phi'$ is the derivative of the potential function with respect to the intermolecular distance r. The assumption of pairwise additivity of the potential energy function is not generally needed in deriving the internal energy and pressure equations. However, definition of the functional form of the intermolecular potential energy function is necessary for these two equations. A relation between the isothermal compressibility, $\kappa_T = -(1/v)(\partial v/\partial P)_T$, and g(r) can be derived without the need to know the functional form of the intermolecular potential. This relation is obtained from the grand canonical ensemble radial distribution function (Hill, 1956)

$$\rho kT\kappa_T = 1 + \rho \int_0^\infty [g(r) - 1] \, 4\pi r^2 \, dr. \tag{21}$$

The corresponding relation for mixtures was derived by Kirkwood and Buff (1951)

$$\rho kT\kappa_T = \frac{|B|}{\rho \sum\sum x_i x_j |B|_{ij}}, \tag{22}$$

where $|B|_{ij}$ symbolizes the cofactor of the element $B_{ij}$ in the c x c matrix B and $|B|$ is the determinant of B. The elements, $B_{ij}$, of the matrix are:

$$B_{ij} = \rho x_i [\delta_{ij} + \rho x_j G_{ij}], \tag{23}$$

where $G_{ij}$ is called the radial distribution function integral, the fluctuation integral, and the Kirkwood-Buff integral. It is defined by the following expression

$$G_{ij} = \int_0^\infty [g_{ij}(r) - 1] 4\pi r^2 \, dr. \tag{24}$$

In addition to the mixture isothermal compressibility, the Kirkwood-Buff solution theory gives the partial molar volumes and the derivatives of the chemical potential with respect to mole fractions in terms of the fluctuation integrals. Details of this theory are discussed in the last section of this chapter.

## DIRECT CORRELATION FUNCTIONS

In the compressibility equation, Equation (21), the term $[g(r)-1]$ in the integrand is known as the total correlation function. This is because it gives the total influence of molecule 1 on molecule 2 at a distance r. Ornstein and Zernike (1914) suggested dividing the total correlation function, $h(r)=g(r)-1$, into two parts, a direct part and an indirect part. They defined the direct correlation function, $c(r)$, by

$$c(r_{12}) = h(r_{12}) - \rho \int c(r_{13}) h(r_{23}) dr_3. \tag{25}$$

The compressibility equation, (21), can be expressed in terms of $c(r)$ by using Fourier transforms (McQuarrie, 1975). The resulting expression is:

$$(kT)^{-1} [\partial P/\partial \rho]_T = 1 - \rho \int c(r) dr. \tag{26}$$

For mixtures the direct correlation function of the pair ij is defined by

$$c_{ij}(r_{12}) = h_{ij}(r_{12}) - \sum_{k=1}^{c} \rho_k \int c_{ik}(r_{13}) h_{jk}(r_{23}) dr_3, \tag{27}$$

where

$$h_{ij}(r) = g_{ij}(r) - 1 \tag{28}$$

and $\rho_k = \rho x_k$.

## DISTRIBUTION FUNCTION APPROXIMATIONS

Obtaining thermodynamic properties from the relations given in the above sections requires knowledge of the distribution functions. Evaluating the distribution functions from microscopic information, such as the intermolecular potential energy, is not easier than evaluating the partition function. However, the distribution functions are easier to deal with because of their physical interpretation. Next we will look into some of the approximations which were suggested to allow calculating the distribution functions without having to perform the large number of integrations in Equations (12) and (13).

In the previous section three basic relations between thermodynamic properties of mixtures and distribution functions were derived. Those relations are the energy equation, (18), the pressure equation, (20), and the compressibility equation, (22). The pressure and energy equations, in the form they are reported here, are based on two assumptions: 1) Classical mechanics and 2) Pairwise additivity of the total intermolecular potential energy. In addition, they are not valid for fluids with angle-dependent potentials in their present form. For such fluids the dependence of the pair distribution function on the various angles needs to be known. The compressibility equation is free from these assumptions. The dependence of the distribution function on the various angles is not necessary for obtaining thermodynamic properties. The fact that the compressibility equation is free from the assumption of any form for the potential energy function is obvious from its derivation. However, its validity for quantum fluids is not so obvious. This was shown by Buff and Brout (1955).

Historically, the first approximation that allows calculating the distribution functions was suggested by Kirkwood. This assumption is called the superposition approximation (Hill, 1956):

$$\rho^{(3)}(r_1,r_2,r_3) = \rho^{(2)}(r_1,r_2)\,\rho^{(2)}(r_1,r_3)\,\rho^{(2)}(r_2,r_3). \tag{29}$$

The motivation for making this approximation is the existence of exact equations (within the pairwise additivity assumption) for n-molecule distribution functions in terms of (n+1)-molecule functions which were first derived by Yvon (1935). For n=2 the equation is (Boublik, 1980):

$$kT\partial\ln\rho^{(2)}(r_1,r_2)/\partial r_1 = -\partial\rho(r_1,r_2)/\partial r_1$$
$$- \int \partial\rho(r_1,r_3)/\partial r_1\, \rho^{(3)}(r_1,r_2,r_3)/\rho^{(2)}(r_1,r_2)\,dr_3. \tag{30}$$

Combining equations (29) and (30) yields the Yvon-Born-Green integro-differential equation (Yvon, 1935; Born and Green, 1946). A similar equation was derived based on the notion of the coupling parameter (Kirkwood and Monroe, 1941; Kirkwood and Boggs, 1942). Unfortunately, both equations are nonlinear and difficult to solve, even numerically. Salpeter (1958) derived another exact relation between the two-body and three-body distribution functions. However, this relation is of limited practical use because of its infinite number of terms which are related to irreducible clusters (Salpeter, 1958).

The method of topological reduction was also utilized to derive integral equations for the distribution functions. In this approach a relation between the total and the direct correlation functions is assumed, in addition to the relation which defines the direct correlation function. The Percus-Yevick (Percus and Yevick, 1958) and the hypernetted chain (van Leeuwen et al., 1959) integral equations are derived using this method. They are based on the assumptions:

$$c(r) = g(r)[1 - e^{\phi/kT}] \quad \text{(Percus-Yevick)} \tag{31}$$

$$c(r) = g(r) - \ln[g(r)] - 1 - \phi/kT \quad \text{(hypernetted chain)} \tag{32}$$

The distribution functions $c(r)$ and $g(r)$ can be solved for by combining either Equations (31) or (32) with the definition of $c(r)$, Equation (27). The extension of these equations to mixtures is straightforward.

Except for the simplest models of pair interactions, such as the hard-sphere model, mixture calculations using the integral equations approach are rather complicated. This problem and the fact that for real fluids the potential energy functions are not generally well understood makes the above approach unsuitable for getting real fluid mixture properties. For the time being, it seems that the best approach is to start from known pure fluid properties (in the form of equations of state, for example) and develop mixture models that utilize these properties in predicting mixture behavior. Some of the models which have been successful in this field are the conformal solution theory and some versions of the perturbation and variational theories (Lucas, 1986; Rowlinson, 1982).

The conformal solution theory (Brown, 1957; Massih and Mansoori, 1983) presents another alternative for calculating mixture properties. The basic assumption in this theory is that all species interact by potential functions which have the same functional forms. They differ only in the values of the potential parameters. Radial distribution functions of mixtures are approximated by that of a pure reference fluid with appropriate scalings of distance, temperature and density. Van der Waals mixture theory is based on the conformal solution theory. A number of other conformal solution theories have been derived. The mean density approximation (Mansoori and Leland, 1972), and the density expansion theory (Mansoori and Ely, 1985a) are among those suggested.

In the perturbation theory (McQuarrie and Katz, 1966; Barker and Henderson, 1967; Weeks, et al., 1971) one divides the intermolecular potential energy function into a reference part and a perturbation part. The reference part represents a potential model for which the thermodynamic properties are known, such as the hard-sphere model. The variational theory (Mansoori and Canfield, 1969; Mansoori and Leland, 1970; Hamad and Mansoori, 1987) provides inequalities which give least upper bound and highest lower bound to the Helmholtz free energy. In both theories a reference system, for which thermodynamic properties and radial distribution functions are known, is needed. Mixture calculations based on these theories, although simpler than the integral equations approach, are still lengthy and usually no closed form expressions can be obtained from them.

## FLUCTUATION THEORY OF MIXTURES

The Kirkwood-Buff solution theory is useful in analysis and prediction of properties of asymmetric and/or highly polar mixtures for which there is limited or no knowledge about their intermolecular potential energy functions (Landau and Lifshitz, 1980). The basic relation of the Kirkwood-Buff solution theory is derived using the grand canonical ensemble theory and it is as follows (Kirkwood and Buff, 1951):

$$kT[\partial <N_i>/\partial \mu_j]_{V,T,\mu_{i \neq j}} = <N_i><N_j>G_{ij}/V + \delta_{ij}<N_i> \tag{33}$$

In this expression k is the Boltzmann constant, T is the absolute temperature, V is the total volume, $<N_i>$ is the average number of particles of type i in the grand canonical ensemble, $\mu_j$ is the chemical potential of component j, $\delta_{ij}$ is the Kronecker delta and $G_{ij}$ is defined by Equation (24). For binary mixtures, Equation (33) reduces to the following equation:

$$[\partial \mu_1 / \partial x_1]_{P,T} = kT / [x_1\{1 + x_1 x_2 \rho (G_{11} + G_{22} - 2G_{12})\}]. \tag{34}$$

Provided information is available about $G_{ij}$ integrals, defined by Equation (24), Equation (34) can be used to calculate chemical potentials of components of a mixture. Also, the information about $G_{ij}$ can provide us with the means of calculating other properties of a mixture such as the isothermal compressibility $\kappa_T$, as given by Equation (22), and the partial molar volumes $v_i$ as given by the following expression:

$$\rho v_i = \sum_{j=1}^{c} x_j |B|_{ij} \Big/ \sum_{j=1}^{c} \sum_{k=1}^{c} x_j x_k |B|_{jk}, \tag{35}$$

where c is the number of components in the mixture, B is a c×c matrix with elements, $B_{ij}$, defined by Equation (23). For binary mixtures the above relation reduce to the following equation:

$$v_1 = \frac{1 + x_2 \rho (G_{22} - G_{12})}{\rho [1 + x_1 x_2 \rho (G_{11} + G_{22} - 2G_{12})]} \tag{36}$$

The above equations have been studied extensively for the case of infinitely dilute solutions (Ben-Naim, 1974 and 1980). The major difficulty in utilizing the fluctuation theory in mixture property calculations is the lack of knowledge about the fluctuation integrals, $G_{ij}$'s. For solutions with finite compositions expansions of the $G_{ij}$'s in powers of concentration are available (Buff and Brout, 1955; Buff and Schindler, 1958). However, the coefficients in these expansions are given in terms of third, fourth and higher order correlation functions. Since little is known about correlation functions of order higher than two, the expansions are in the time being of limited practical use.

The Kirkwood-Buff fluctuation theory has been used by Mazo (1958) as a starting point for the development of a theory based on the perturbation expansion of the excess free energy of mixtures. It has been applied to classical mixtures as well as to quantum mechanical isotope mixtures.

Theoretical calculation of $G_{ij}$ integrals requires the knowledge about the radial distribution functions, $g_{ij}$. The radial distribution functions can generally be calculated using the theory of intermolecular potential energy functions in the context of a partition function. However, for complex mixtures, such as mixtures consisting of asymmetric, highly polar, and associating molecules the intermolecular potential energy functions are not very well known. Also, the existing techniques of calculating radial distribution functions from the knowledge of intermolecular potential energy functions require extensive numerical calculations (McQuarrie, 1975).

Pearson and Rushbrooke (1957) reformulated the fluctuation theory expressions, Equations 34-36 in terms of direct correlation function integrals, $C_{ij}$, as defined by

$$C_{ij} = \int_V c_{ij}(\mathbf{r}) d\mathbf{r}. \tag{37}$$

In this equation $c_{ij}(\mathbf{r})$ is the direct correlation function, which is related to the radial distribution function as defined by Equations (27) and (28). The chemical potential is related to the direct correlation function integrals as follows (O'Connell, 1971):

$$(kT)^{-1} [\partial \mu_i / \partial \rho_j]_{T, \rho_{k \neq j}} = \delta_{ij}/\rho_i - C_{ij} \tag{38}$$

For example, for binary mixtures the relation between the chemical potential and the direct correlation function integrals will reduce to the following equation:

$$x_1[\partial\mu_1/\partial x_1]_{P,T} = kT \frac{1-x_1\rho C_{11}-x_2\rho C_{22}+x_1x_2\rho^2(C_{11}C_{22}-C_{12}^2)}{1-\Sigma\Sigma x_i x_j \rho C_{ij}} \tag{39}$$

In terms of the direct correlation function integrals the isothermal compressibility and the the partial molar volume become (Pearson and Rushbrooke, 1957; O'Connell, 1971).

$$(\rho kT\kappa)^{-1} = 1-\rho\sum_{i=1}^{c}\sum_{j=1}^{c} x_i x_j C_{ij} \tag{40}$$

and

$$\rho \bar{v}_i = \left(1-\rho\sum_{j=1}^{c} x_j C_{ij}\right) \bigg/ \left(1-\rho\sum_{j=1}^{c}\sum_{k=1}^{c} x_j x_k C_{jk}\right) \tag{41}$$

According to Equations (38)-(41) the Kirkwood-Buff expression is reformulated with respect to another molecular function, $c_{ij}$ rather than $g_{ij}$. It should be pointed out that the following general relation in matrix form holds between the fluctuation integrals, $G_{ij}$'s, and the direct correlation function integrals, $C_{ij}$'s (O'Connell, 1971),

$$\underline{\underline{G}} = \underline{\underline{C}} + \underline{\underline{G}}\, \underline{\underline{X}}\, \underline{\underline{C}} \tag{42}$$

where $\underline{\underline{G}}$ is a cxc matrix with elements $G_{ij}$ as given by Eq. (24), $\underline{\underline{C}}$ is a cxc matrix with elements $C_{ij}$ given by Eq. (37), and $\underline{\underline{X}}$ is a diagonal matrix with elements $X_{ij}=<N_i/N>$. Eq. (42) reduces to the following expressions in a binary mixture

$$\rho G_{ii} = \frac{\rho C_{ii} - (1-x_i)\rho^2(C_{11}C_{22}-C_{12}^2)}{1-x_1\rho C_{11}-x_2\rho C_{22}+x_1x_2\rho^2(C_{11}C_{22}-C_{12}^2)}, \quad i=1,2 \tag{43}$$

$$\rho G_{12} = \frac{\rho C_{12}}{1-x_1\rho C_{11}-x_2\rho C_{22}+x_1x_2\rho^2(C_{11}C_{22}-C_{12}^2)}, \tag{44}$$

or

$$\rho C_{ii} = \frac{\rho G_{ii} + (1-x_i)\rho^2(G_{11}G_{22}-G_{12}^2)}{1+x_1\rho G_{11}+x_2\rho G_{22}+x_1x_2\rho^2(G_{11}G_{22}-G_{12}^2)}, \quad i=1,2 \tag{45}$$

$$\rho C_{12} = \frac{\rho G_{12}}{1+x_1\rho G_{11}+x_2\rho G_{22}+x_1x_2\rho^2(G_{11}G_{22}-G_{12}^2)} . \tag{46}$$

The equations and conceptions illustrated in the previous sections constitute a useful basis for a better comprehension of the various subjects presented in the following chapters of this volume. The books and articles quoted in the present paper are the most appropriate for those readers who look for a deeper mastering of the matter.

ACKNOWLEDGMENT. This work has been supported by the NATO Advanced Study Program.